\documentclass[reprint,twocolumn,preprintnumbers,pre,amsmath,superscriptaddress,amssymb]{revtex4}
\usepackage{dcolumn}
\usepackage{hyperref}
\usepackage{color}
\usepackage{indentfirst}
\usepackage{graphicx}
\usepackage{bm}
\usepackage{float}

\usepackage{amsmath}
\begin{document}
\title{Optical Thouless conductance and level-spacing statistics in two-dimensional Anderson localizing systems}
\author{Sandip Mondal}
\affiliation{Nano-optics and Mesoscopic Optics Laboratory, Tata Institute of Fundamental Research, 1, Homi Bhabha Road, Mumbai, 400 005, India}
\author{Randhir Kumar}
\affiliation{Nano-optics and Mesoscopic Optics Laboratory, Tata Institute of Fundamental Research, 1, Homi Bhabha Road, Mumbai, 400 005, India}\author{Martin Kamp}
\affiliation{Lehrstuhl f\"{u}r Technische Physik, Universit\"{a}t W\"{u}rzburg, 97074 W\"{u}rzburg, Germany.}
\author{Sushil Mujumdar}
\email[]{mujumdar@tifr.res.in}
\homepage[]{http://www.tifr.res.in/\~mujumdar}
\affiliation{Nano-optics and Mesoscopic Optics Laboratory, Tata Institute of Fundamental Research, 1, Homi Bhabha Road, Mumbai, 400 005, India}
\date{\today}
\begin{abstract} We experimentally investigate spectral statistics in Anderson localization in two-dimensional amorphous disordered media. Intensity distributions captured over an ultrabroad wavelength range of $\sim 600$~nm and averaged over numerous configurations provided the Ioffe-Regel parameter to be $\sim2.5$ over the investigated wavelength range. The spectra of the disordered structures provided access to several quasimodes, whose widths and separations allowed to directly estimate the optical Thouless conductance $g_{Th}$, consistently observed to be below unity. The probability distribution of $g_{Th}$ was measured to be a log-normal. Despite being in the Anderson localization regime, the spacings of energy levels of the system was seen to follow a near Wigner-Dyson function. Theoretical calculations based on the tight-binding model, modified to include coupling to a bath, yielded results that were in excellent agreement with experiments. From the model, the level-spacing behavior was attributed to the degree of localization obtained in the optical disordered system.
\end{abstract}
\maketitle

Anderson localization is an interesting transport phenomenon in disordered systems, first proposed in 1958 for electronic systems \cite{sheng,Akkermans,Anderson58}. In presence of impurities in metal, the diffusive motion of electrons is completely arrested due to self-interference of the multiple scattered electron waves. Being a general wave phenomenon, this concept has immediately percolated to other areas of physics such as, photonics, acoustics, matter waves, etc \cite{Segev13,Wiersma13,Hu08,Billy08,kondov11}. Among these, photon localization has triggered immense research due to light-specific advances such as quantum\cite{Gilead15,Crespi13,giuseppe13,Lahini10} and nonlinear\cite{Schwartz07,Lahini08,Levi12} transport, localization-based lasing\cite{cao99,genack_loc05,liu14,randhir17,mafi17}, vectorial scattering\cite{skipetrov14} and so on. Structural correlations have been reported to realize new effects in transport such as bandgap formation in the absence of translational order\cite{florescu09,froufe-perez16}, novel transition from localization to bandgap domain\cite{froufe-perez17}, and modification of the lcoalization length over orders of magnitude\cite{conley14}. The achievement of unambiguous three-dimensional localization is challenging\cite{Wiersma97,sperling13, sperling16,Storzer06}, due to which lower-dimensional structures have been used to investigate the rich physics of disorder\cite{Schwartz07,nahata17,sapienza10,garcia12,lodahl12,Szameit10}. Furthermore, low-dimensional systems also allow for direct access to the exponential wavefunction which conclusively characterizes Anderson localization in the absence of loss.

Similar to the wavefunction, another feature that characterizes localization is the spectrum of disorder that reflects the energy-levels of the structure. Transport in a disordered system occurs via the formation of multiple resonances situated at random locations in space and frequency, associated with random widths originating from their coupling to the bath. These quasimodes constitute the disorder spectrum. Diffusive transport occurs when the spectral widths of the quasimodes are larger than their separations enhancing inter-quasimode coupling. Under strong disorder, the widths are smaller, inhibiting the intermode coupling, and the system transits into the localization domain. Accordingly,  the domain of transport is characterized by a spectral parameter called the Thouless conductance, defined as $g_{Th}=\delta \omega/\Delta \omega = \delta \lambda/\Delta \lambda$, where $\delta \omega (\delta \lambda)$ is the average spectral width of two adjacent modes and $\Delta \omega (\Delta \lambda)$ is the separation between the modes\cite{thouless77}. Furthermore, another spectral effect arises under disorder. In periodic systems, the energy levels are correlated across the spectrum. As disorder is introduced, the correlations fall, and under conditions of localization, the eigenvalues ($\omega$'s) are expected to be completely uncorrelated. This is reflected in the statistics of the spacings between consecutive levels $s = \omega_i - \omega_{i-1}$, where localizing systems exhibit Poissonian spacings, while diffusive systems show a spacing distribution approximated by the Wigner Dyson function given as $\sim (\pi s/2) \exp{(-\pi s^2 /4)}$\cite{Izrailev90}. While the theoretical aspects of the spectra of disorder have been available in literature, to our knowledge, there are no experimental reports which verify the same. The primary challenge therein is the requirement of ultrabroad spectral range for measuring the disorder spectrum, and a sufficiently large ensemble of configurations for the statistics. In this communication, we achieve precisely the same, by employing samples with 75 configurations of amorphous disorder, whose spectra were measured over a range of 600 nm. Over this range, we access about 30 localized quasimodes in each configuration. The localization length is quantified from the configurationally-averaged intensity distributions. The measured spectra allow for the extraction of the quasimodes and quantification of the optical Thouless conductance. Subsequently, the level spacing statistics are measured from the same spectra, and exhibit a close correspondence with the Wigner-Dyson function despite being in the localization domain. Theoretical computations were carried out based on the tight-binding Hamiltonian, whose eigenvalues were further subjected to broadening due to coupling to the bath.  The computational results are in excellent agreement with the experiments, and relate the level-spacing statistics to the degree of localization in optical systems.

\begin{figure}[htbp]
	\centering
	\includegraphics[width=8.0cm]{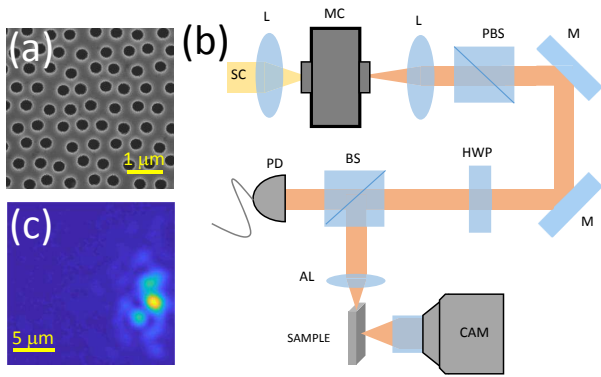}
	\caption{(a) SEM image of part of a disordered sample. (b) Experimental setup. Legend: SC: Supercontinuum, L: Lens, MC: Monochromator, PBS: Polarizing Beam Splitter, M: Mirror, HWP: Half Wave plate, BS: Beam Splitter, PD: Photo Detector, AL: Aspheric Lens, CAM: IR Camera.(c) Measured mode profile at a particular wavelength in one configuration.}
	\label{fig:Expt_setup}
\end{figure}

For the experiments, disordered structures are fabricated in Gallium Arsenide membranes (thickness 340~nm). Air holes (radius 139~nm) are lithographically written on the membrane at pre-defined disorder sites. To avoid band-tail localization, amorphous disorder was realized, and was confirmed by the Fourier transform of the refractive index distribution. Further, the calculation of the Structure factor $S({\bf q})$ for these configurations shows that there are no structural correlations in the wavelength range of our interest\cite{supp1}. The dimensions of the structured sample are 20~$\mu$m $\times$ 20~$\mu$m. For statistical completeness, experiments are carried out over 75 configurations.

Figure \ref{fig:Expt_setup}(a) shows the scanning electron micrograph of a section of a representative configuration. (b) depicts the experimental setup showing a broadband IR beam  of a supercontinuum source (Fianium, CW power 4W, $\lambda = 1050$~nm to $1650$~nm). The beam is passed through a monochromator to obtain a tunable narrowband light with a spectral width $\sim2-3~nm$, which is sufficiently narrow to excite individual modes. Next, the beam is passed through a combination of a polarizing beam splitter and a half-wave plate to achieve the desired input polarization. A $90:10$ beamsplitter allows 10\% of the beam to be incident onto a photodetector, which monitors the power incoupled into the sample. The rest of the beam is focused by an aspheric lens onto the edge of the sample. The incoupled light excites the available modes, which are mapped by measuring the out-of-plane scattered light. The scattered light is imaged by a SWIR (Short Wavelength Infra Red) camera aided by a 100X objective. Intensity profiles are recorded over the entire range of the wavelengths (1050~nm-1650~nm), in steps of 2~nm. Figure \ref{fig:Expt_setup}(c) illustrates a measured mode at a particular wavelength and configuration. A localized mode in the vicinity of the input edge is identified readily\cite{supp2}.  The strong disorder in the structures does not allow the light to propagate deeper in the system. The localized character was reconfirmed via intensity statistics $P(I/\langle I \rangle)$, which exhibited a long-tailed deviation from exponential statistics, allowing us to estimate the dimensionless conductance $g$\cite{rossum99,supp3}.

\begin{figure}[htbp]
	\centering
	\includegraphics[width=8.0cm]{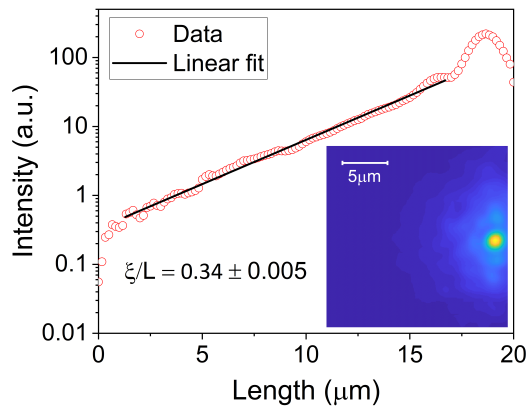}
	\caption{Exponential tail of the Anderson localized modes, obtained from the cross-section of the configurationally-averaged intensity shown in the inset. Here, $\lambda = 1110$~nm.}
	\label{fig:loc_length}
\end{figure}

Figure~\ref{fig:loc_length} depicts the measurement of the localization length in the structures. The inset shows a configurationally-averaged intensity distribution at a representative $\lambda = 1110$~nm. The main plot shows a cross-section (on a logarithmic Y axis) through the intensity maximum. The tail shows a clear exponential decay, which was characterized to yield $\xi/L=0.34$, where $L = 20~\mu$m, the sample dimension. The loss length due to the out-of-plane scatter was calibrated to be $\sim5L$, which is substantially larger than the measured $\xi$\cite{supp4}. Similar characterization of $\xi$ was carried out over all wavelengths.

\begin{figure}[htbp]
	\centering
	\includegraphics[width=8.0cm]{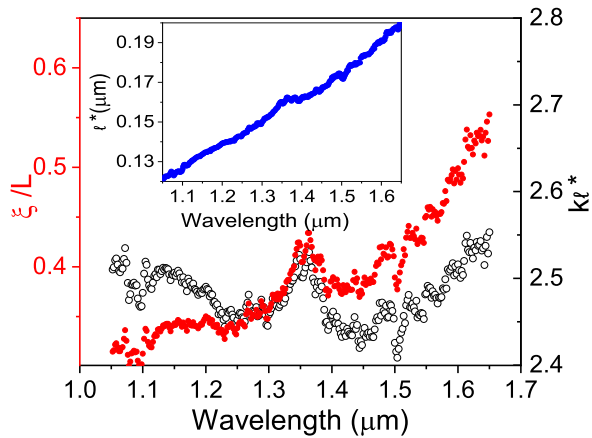}
	\caption{(a) Localization length $\xi$ (red dots, left Y-axis), and $k\ell^*$ (black circles, right Y-axis) as a function of wavelength. Inset: $\ell^*$ as extracted from $\xi$.}
	\label{fig:klstar_lstar}
\end{figure}

Figure~\ref{fig:klstar_lstar} shows the variation of $\xi/L$ (red dots) with wavelength. A gradually increasing profile is observed, with the $\xi$ ranging from 0.3L to 0.55L. The increase is related to the scattering cross-section of the individual scatterer (air hole), where larger wavelengths experience weaker scattering and hence a larger $\xi$. This fact is reflected in the inset, which shows the $\ell^*$ as a function of $\lambda$. The $\ell^*$ is extracted from the expression of $\xi$ in two-dimensions $\xi_{loc}=\ell^*exp(\pi k \ell^*/2)$. The $\ell^*$ is much smaller than the operating range of wavelengths.
A kink is noted at $\lambda\sim1350$~nm, the origin of which is unclear at this stage. It is also existent in the $\ell^*$. The black circles show the spectral variation of the Ioffe-Regel parameter $k\ell^*$, which is range-bound between 2.4 and 2.55. The fact that $k\ell^*\nleqslant 1$ indicates that the modes are not very tightly localized, which has a bearing on the level-spacings as discussed later.

\begin{figure}[htbp]
	\centering
	\includegraphics[width=8.0cm]{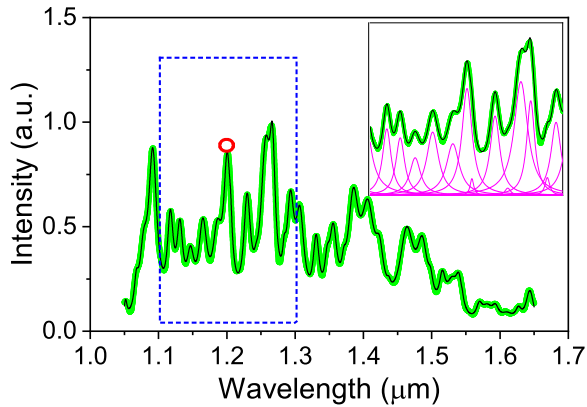}
	\caption{ Intensity (green markers) as a function of $\lambda$. Black line is a fit to the data using a sum of Lorentzians. Red circle corresponds to the intensity image in Fig~\ref{fig:Expt_setup}(c). Inset shows the boxed region, with the ingredient Lorentzians explicitly shown.}
	\label{fig:Spectrum}
\end{figure}
Energy spectra were then constructed by choosing a spatial position (\textbf{x},\textbf{y}) in the region of the mode, and picking the recorded intensity $I(\textbf{x},\textbf{y},\lambda)$\cite{Riboli}. A representative spectrum (green dots) is shown in Fig.~\ref{fig:Spectrum}. The peaks in the spectrum indicate the resonant modes of the system. The red circle corresponds to the intensity distribution shown in Fig~\ref{fig:Expt_setup}(c). To isolate the resonances, a sum of Lorentzians is used to fit the spectrum, where the peak amplitudes, positions and the widths of the Lorentzians are fit parameters\cite{sebbah06}. The black line in the plot is the fit spectrum. A section of the spectrum (marked by the dashed rectangle) is emphasized in the inset, exhibiting the various participating Lorentzians\cite{supp5}.

\begin{figure}[htbp]
	\centering
	\includegraphics[width=8.0cm]{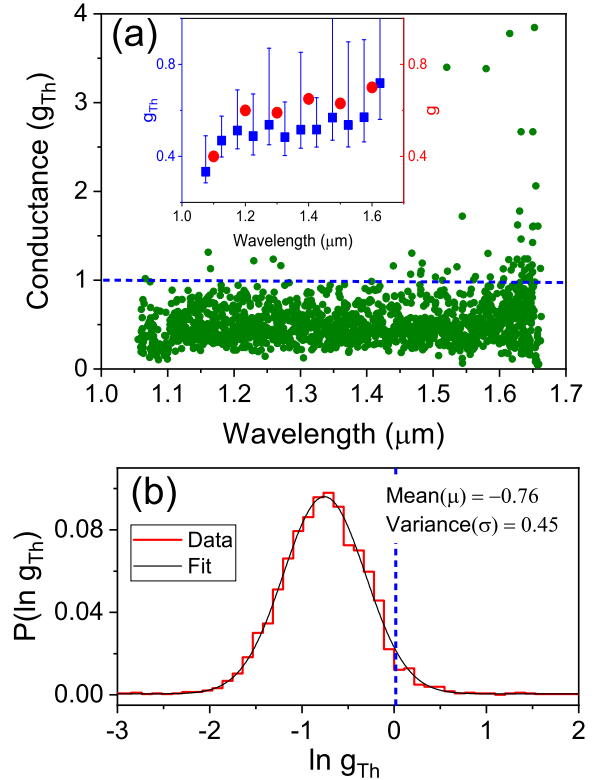}
	\caption{(a) Scatter-plot of measured Thouless conductance $g_{Th}$ over all wavelengths and configurations. Blue dashed line separates localized ($g_{Th}<1$) and delocalized ($g_{Th}>1$) modes. Inset: Spectral variation of $g_{Th}$ (blue squares, left Y-axis), where the markers represent $\langle g_{Th} \rangle$ and error bars signify the standard deviation. Red circles (right Y-axis) show $g$ computed from the $P(I/\langle I \rangle)$. (b) Measured $P(\ln~g_{Th})$, showing a Gaussian distribution as is theoretically predicted for localized modes. Black line is the Gaussian fit. Blue dashed line separates localized and delocalized modes.}
	\label{fig:Conductance}
\end{figure}

Next, the Thouless conductance $g_{Th}$ is calculated from the Lorentzian widths and separation between the Lorentzians as extracted from the fit routine. When $g_{Th}>1$, the modes overlap spectrally and the system transports light through intermode energy transfer. However, if $g_{Th}<1$, the Thouless criterion for Anderson localization is satisfied. Figure \ref{fig:Conductance}(a) shows the scatter plot of the $g_{Th}$ over all configurations. The blue dashed line placed at $g_{Th}=1$ separates the localized and delocalized modes. Clearly, a vast majority of the modes are localized. Few configurations exhibit extremely tight localization with $g_{Th} \rightarrow 0.1$. A major part of the scatterplot is flat, revealing a spectral insensitivity of $g_{Th}$. However, several outliers are seen with $g_{Th}>1$, particularly at larger $\lambda$, where some configurations show strong delocalization with $g_{Th}>3.5$. The outliers induce a wavelength-dependence in $g_{Th}$, shown in the inset (blue squares, left Y-axis) where each $\lambda$-bin is of 50~nm.  The asymmetric error-bars represent the asymmetry in $P(g_{Th})$. Clearly, the $\langle g_{Th} \rangle$ rises with $\lambda$. The inset also depicts the conductance $g$ (red circles, right Y-axis) calculated from the $P(I/\langle I \rangle)$. It can be seen that $g \gtrsim g_{Th}$ over the displayed energy range.  Finally, the distribution of $\ln g_{Th}$ is shown in Figure~\ref{fig:Conductance}(b). The red curve shows the experimental histogram, which is a perfect Gaussian. As is well-understood in the literature, the conductance is seen to be log-normally distributed\cite{rossum99,rotter17}. The black curve is the fit to the data, and reveals a $\langle \ln g_{Th} \rangle = -0.76$, and a width of 0.45. These are, to our knowledge, the first direct measurements of optical Thouless conductance and its distribution in two-dimensional, optical Anderson localizing media.

The log-normal nature of the distribution endorses the strong localization of the modes. However, the $\xi$ indicates that there is further scope for tighter localization. This discrepancy arises from the inherent structure. Earlier experiments in two-dimensional membranes has shown that, when the number of air-holes is increased to augment the disorder, the scattering loss also increases and so does the width of the Lorentzians\cite{Riboli}. This effectively weakens the localization of the modes.

\begin{figure}[htbp]
	\centering
	\includegraphics[width=8.0cm]{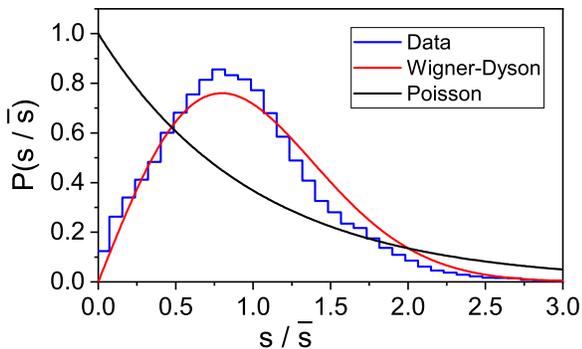}
	\caption{Blue line shows the experimentally measured distribution of energy level spacings of the disordered samples. Red curve depicts the Wigner-Dyson distribution, while the black line indicates Poissonian distribution. The measured modes exhibit level repulsion despite their localized character.}
	\label{fig:level_expt}
\end{figure}

\begin{figure}[htbp]
	\centering
	\includegraphics[width=8.0cm]{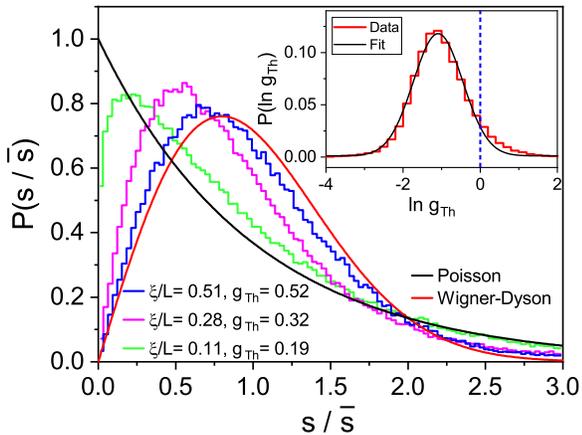}
	\caption{Computed $P(s/\bar{s})$ from a tight-binding Hamiltonian. Red and black curves depict the Wigner-Dyson and Poissonian distributions respectively. The histograms indicate $P(s/\bar{s})$ for varying degree of localization as indicated in the legend. $P(s/\bar{s})$ is closer to a Wigner-Dyson distribution, even for localized modes for the experimentally achieved degree of localization. Inset shows the $P(\ln g_{Th})$, for $\xi/L = 0.51$.}
	\label{fig:level_th}
\end{figure}

The procedure for measuring the Thouless conductance directly provides access to the eigenfrequencies $\omega_i$ of the disorder, which enables us to investigate the level-spacing statistics. The spacing $s = \omega_i - \omega_{i-1}$ is an interesting parameter that characterizes random spectra of disordered systems. Figure~\ref{fig:level_expt} depicts (blue curve) the histogram of $s/\bar{s}$, the spacings normalized to mean spacing. Clearly, $P(s/\bar{s}) \rightarrow 0$ as $s/\bar{s} \rightarrow 0$, which indicates a mode repulsion. This arises from inherent correlations in the eigenfunctions. The red profile illustrates the Wigner-Dyson (WD) function. The black curve shows the Poisson distribution, describing spacings between completely uncorrelated eigenfunctions. The $P(s/\bar{s})$ behaves almost congruent to WD, rising linearly, maximising close to 1, and decaying with a Gaussian tail. Theoretically, localized eigenfrequencies are expected to be uncorrelated, and hence $P(s/\bar{s})$ ought to be Poissonian. However, in realistic finite-sized systems, the localization is not too tight, and deviations towards WD are expected. For instance, in a recent computational study on disordered photonic crystals\cite{Escalante18}, it was shown that the $P(s/\bar{s})$ remains similar to WD function despite the system entering localization domain. In order to support these observations, we implemented a disorder model in the form of a tight-binding Hamiltonian with diagonal disorder, $H=\sum_{i} [ \Omega_{i} c^{\dagger}_ic_i +  \sum_{j} p c^{\dagger}_i  c_{i+j} +  H.c.]$, where the $c_i$ is the annihilation operator, ‘$p$’ is the hopping probability between the sites, and $j$ runs over the nearest neighbours. The diagonal term is determined by a uniform random variate $\Omega_{i} \in [1 - W, 1 +W]$ where $W$ varies from 0.1 for weak disorder to 1 for very strong disorder. The hopping probability ‘$p$’ is kept constant at 0.1. The Hamiltonian matrix is diagonalized to find the eigenvalues $\omega_i$ and eigenvectors $\psi_i$ of the disordered system. In postprocessing of data, the eigenvalues were broadened  (giving a width $\delta \omega_i$) by a loss factor calculated as $\Gamma \propto \exp(-2R/\xi)$, where $R$ quantified the distance of the peak of the $|\psi_i|^2$, averaged over the four boundaries\cite{Sebbah07}. One thousand configurations are computed for statistical averaging. Accordingly, the computation provides both $g_{Th} = \delta \omega / \Delta \omega$ and $P(s/\bar{s})$. Figure~\ref{fig:level_th} shows the $P(s/\bar{s})$ for three disorder strengths, and the legend mentions corresponding values of $\langle g_{Th} \rangle$ and $\xi/L$. The level-spacing distribution tends to a Poissonian with increasing disorder. Clearly, for the magnitude of conductance obtained in our experiments, the $P(s/\bar{s})$ is still close to the Wigner-Dyson function. The inset shows the computed $P(\ln g_{Th})$, endorsing the log-normal distribution of $g_{Th}$.

In conclusion, we designed and fabricated amorphous disordered templates in GaAs membranes. Measuring the intensity profiles over an ultrabroad wavelength range and numerous configurations, we extracted the optical Thouless conductance and the distribution $P(\ln g_{Th})$ thereof. For comparison, the dimensionless conductance $g$ was extracted from $P(I/\langle I \rangle)$, and was found to be very comparable, albeit slightly larger than $g_{Th}$. The measured $P(\ln g_{Th})$ exhibited a Gaussian distribution, consistent with the Anderson localized domain of transport, as was independently verified.  The level spacing statistics were experimentally measured, and suggested a likeness to the Wigner Dyson function despite the localized transport, which typically shows a Poissonian behavior. The origin of this observation was the moderate degree of localization obtained in this amorphous system. We implemented a tight-binding Hamiltonian with a strong disorder in the nearest-neighbour couplings, with a loss parameter added in the postprocessing of eigenvalues. The model excellently reproduced the $P(\ln g_{Th})$ and $P(s/\bar{s})$ observed in the experiments for a comparable degree of localization.
These observations are generally representative of the behavior of light in localizing systems, and indicative of the disorder strength that such systems can offer. Riboli et al have demonstrated the inefficiency of increasing disorder by raising either the hole density or the hole radius\cite{Riboli}. Our sample sizes are also typical of the large-area sizes in membranes. Coupled to these results, therefore, one can infer that the Wigner-Dyson may turn out to be the limiting distribution for level statistics in practical optical systems. We believe that these studies shed important light on localization in optical systems, a research area which is already seeing rapid novel developments.

\section{Acknowledgements} We are thankful to Kevin Vynck for very helpful discussions on Structure factor calculations. We acknowledge funding from Department of Atomic Energy, Government of India, Plan Proposal of TIFR No XII-P0243, and DST-DAAD Personnel Exchange Programme. We acknowledge expert sample fabrication by Monika Emmerling. SM acknowledges the Swarnajayanti Fellowship from the Department of Science and Technology, Government of India.

\end{document}